\documentclass[%
 reprint,
 amsmath,amssymb,
 aps,
]{revtex4-1}

\usepackage{graphicx}
\usepackage{dcolumn}
\usepackage{bm}
\usepackage{color}

\begin{document}

\preprint{APS/123-QED}

\title{Interplay of activation kinetics and the derivative conductance determines the resonance properties of neurons}

\author{Rodrigo F.O. Pena$^1$}
\author{Cesar C. Ceballos$^{1,2}$}%
\author{Vinicius Lima$^1$}
\author{Antonio C. Roque$^1$}
\email{antonior@usp.br}
\affiliation{%
 $^1$Department of Physics, School of Philosophy, Sciences and Letters of Ribeir\~ao Preto, University of S\~ao Paulo, Ribeir\~ao Preto, Brazil. $^2$Department of Physiology, School of Medicine of Ribeir\~ao Preto, University of S\~ao Paulo, Ribeir\~ao Preto, Brazil
}%




\date{\today}

\begin{abstract}
In a neuron with hyperpolarization activated current ($I_h$), the correct input frequency leads to an enhancement of the output response. This behavior is known as resonance and is well described by the neuronal impedance. In a simple neuron model we derive equations for the neuron's resonance and we link its frequency and existence with the biophysical properties of $I_h$. For a small voltage change, the component of the ratio of current change to voltage change ($dI/dV$) due to the voltage-dependent conductance change ($dg/dV$) is known as derivative conductance ($G_h^{Der}$). We show that both $G_h^{Der}$ and the current activation kinetics (characterized by the activation time constant $\tau_{h}$) are mainly responsible for controlling the frequency and existence of resonance. The increment of both factors ($G_h^{Der}$ and $\tau_{h}$) greatly contributes to the appearance of resonance. We also demonstrate that resonance is voltage dependent due to the voltage dependence of $G_h^{Der}$. Our results have important implications and can be used to predict and explain resonance properties of neurons with the $I_h$ current.
\end{abstract}

\pacs{Valid PACS appear here}
\maketitle


\section{\label{sec:introduction}Introduction}
When stimulated with oscillatory inputs, some neurons respond in preferential frequencies. This resonance phenomenon is demonstrated by an enhancement of the output amplitude. The neuronal impedance is the measure normally used to identify resonance \cite{hutcheon2000}. It shows how much of the input frequency is contained into the output. When resonance is present, a prominent peak is identified at the impedance profile. 

Generally speaking, the electrical properties of passive membranes can be represented by an equivalent RC circuit \cite{gerstner2014}. Leak currents, which are ideally instantaneous and non voltage-dependent, are described by electrical resistances between intracellular and extracellular media, and capacitors describe the charge separation between the two sides of the bilipid membrane due to different ion concentrations on the two sides. This type of electrical circuit works as a low pass filter where an increase in the frequency of input leads to a decrease in the output voltage. 

In addition to this simple circuit, most neurons also express voltage-dependent ion channels that carry voltage-dependent currents. The presence of voltage-dependent currents drastically changes the equivalent electrical circuit, and in some cases make it work as a bandpass filter where a resonance peak arises. It is well known that the impedance profile at subthreshold voltages is mainly determined by the hyperpolarization activated current ($I_h$) in several neuron types \cite{Ge2016}, e.g. the pyramidal cells of the hippocampus \cite{dougherty2013,Zemankovics2010}. The $I_h$ current has been called a ``resonant current'' elsewhere \cite{izhikevich2007}. However, the biophysical mechanisms underlying the resonance generation by $I_h$ in neurons remain unclear.

$I_h$ is a slowly non-inactivating current with an activation time constant ($\tau_h$) that spans a range from tens of milliseconds to several seconds \cite{dougherty2013,ceballos2016,surges2003,poolos2006,vanwelie2004,Zemankovics2010}. $I_h$ is a voltage-dependent current and neurons with $I_h$ display an impedance magnitude that is also voltage-dependent \cite{Zemankovics2010}. Interestingly, the simple expression of $I_h$ in a neuron's membrane is not sufficient to cause resonance. For instance, it has been observed that in the presence of $I_h$ there is no resonance for membrane potentials too depolarized or too hyperpolarized, or when $\tau_h$ is too small \cite{rotstein2014, hutcheon1996a, hutcheon1996b}. Furthermore, the resonance frequency also varies in a voltage-dependent manner \cite{Zemankovics2010}. However, the source of this voltage dependency has not yet been identified.

The $I_h$ current can be expressed as the product of a conductance by a driving force, $I_h = g(V,t)(V - E)$, where $g(V,t)$ is the so-called chord conductance and $E$ is the reversal potential \cite{gerstner2014,izhikevich2007}. Thus, for small voltage changes the variation of $I_h$ with respect to $V$ is $dI_h/dV = g + (dg/dV)(V - E)$, where the second term is the so-called derivative conductance ($G_h^{Der}$) \cite{ceballos2017review}. While $g$ reflects the passive changes of the current, $G_h^{Der}$ reflects the changes in the current due to voltage-dependent conductance changes ($dg/dV$). Both conductances, $g$ and $G_h^{Der}$, are voltage-dependent and contribute to the impedance magnitude and the generation of neuronal resonance \cite{rotstein2014}. However, it is still unknown the relative contribution of each conductance to the resonance mediated by $I_h$.  

Experimentally, it is known that $I_h$ attenuates slow neuronal voltage changes, acting as a high-pass filter. This attenuation reduces the impedance magnitude at low frequencies. The strength of the attenuation is directly proportional to both $I_h$ conductances (chord and derivative) and inversely proportional to the $I_h$ activation kinetics \cite{rotstein2014}.

The main goal of this paper is to determine the mechanisms underlying the resonance induced by $I_h$ in a simple neuron model containing only leak and $I_h$ currents. In our simulations we used biophysical parameters to reproduce the impedance properties of CA1 pyramidal cells of the hippocampus. This neuron displays resonance due to the $I_h$ current and its time constant is better fitted by the sum of two exponentials, namely the fast and the slow time constants \cite{dougherty2013,Zemankovics2010}. Whereas the fast component has values close to tens of milliseconds, the slow component has values from hundreds of milliseconds to approximately one second \cite{surges2003,dougherty2013,poolos2006,vanwelie2004}.

We ask how the range of $\tau_h$ values contribute to the impedance profiles, the existence of resonance, and the values of the resonance frequency. We determine the voltage-dependent impedance profiles while changing the values of both leak and $I_h$ conductances as well as of $\tau_h$. Our results show that the derivative conductance and $\tau_h$ are the main factors in the generation of resonance in the simulated neuron. 

\section{\label{sec:methods}Methods}

\subsection{\label{sec:neuron_model}Neuron Model}

In our neuron model, we consider a single compartment where the membrane has its voltage described by 
\begin{equation}
C \frac{dV}{dt}= - I_{h} - I_{L} + I(t),
\label{Eq:model}
\end{equation}

\noindent where $C$ is the membrane capacitance, $I_L$ is a leak current, $I_h$ is the hyperpolarization activated current, and $I(t)$ is an external current. The $I_h$ current is modeled using the Hodgkin-Huxley formalism obeying 
\begin{equation}
I_{h} = \bar{g}_h A_h(V,t)(V-E_h),
\label{Eq:h_current}
\end{equation}

\noindent with maximum conductance $\bar{g}_h$ in units of nS and reversal potential $E_h=-30$ mV. 

The activation variable $A_h$ is represented as
\begin{equation}
\frac{dA_h(V,t)}{dt} = \frac{A^{\infty}_h(V)-A_h(V,t)}{\tau_h},
\label{Eq:Ah_kinetics}
\end{equation}

\noindent where $\tau_h$ is the activation time constant in units of ms and $A^{\infty}_h$ is the steady state activation variable. $A^{\infty}_h$ is voltage-dependent and obeys the Boltzmann function
\begin{equation}
A^{\infty}_h = \frac{1}{1 + \exp\left(\frac{V-V_{1/2}}{k}\right)},
\label{Eq:Ainf}
\end{equation}

\noindent where $V_{1/2}=-82$ mV and $k=9$ mV. Observe that $V_{1/2}$ represents the voltage in which $A^{\infty}_h=0.5$ and $k$ is the slope of the $A^{\infty}_h$. Both $V_{1/2}$ and $k$ are fitted experimentally \cite{magee1998}.

The leak current is modeled following $I_L= g_L(V- E_L)$ where $g_L$ is the maximum conductance in units of nS and the reversal potential $E_L=-90$ mV. The model parameters are within the physiological range for a CA1 pyramidal cell in the hippocampus \cite{ceballos2017}.

\subsection{\label{sec:slope_chord}Slope, derivative and chord conductance}

The $I_h$ slope conductance ($G_h$), i.e. the slope of the steady-state IV plot, of our model is obtained by differentiating Eq.~(\ref{Eq:h_current}) with respect to $V$,
\begin{equation}
G_h = \frac{dI_h}{dV} = \underbrace{\bar{g}_h A^{\infty}_h}_{\text{{chord}}} + \underbrace{\bar{g}_h (V-E_h) \frac{dA^{\infty}_h}{dV}}_{\text{{derivative}}} ,
\label{Eq:slope}
\end{equation}

\noindent where the first term is the chord conductance ($g_h$) and the second term is the derivative conductance ($G_{h}^{Der}$) \cite{ceballos2017}. More specifically, the derivative of $A^{\infty}_h(V)$ in Eq.~(\ref{Eq:slope}) can be obtained differentiating Eq.~(\ref{Eq:Ainf}) leading to $\frac{dA^{\infty}_h}{dV}=\frac{(A^{\infty}_h - 1)A^{\infty}_h}{k}$. In the end, we can write the derivative conductance as: 
\begin{equation}
G_{h}^{Der} = \bar{g}_h A^{\infty}_h \frac{\left(A^{\infty}_h - 1\right)}{k}(V-E_h).
\label{Eq:slope_detailed}
\end{equation}

For the case of $I_h$, both the chord conductance and the derivative conductance are positive. 

\begin{figure}[!ht]
\includegraphics[width=0.4\textwidth]{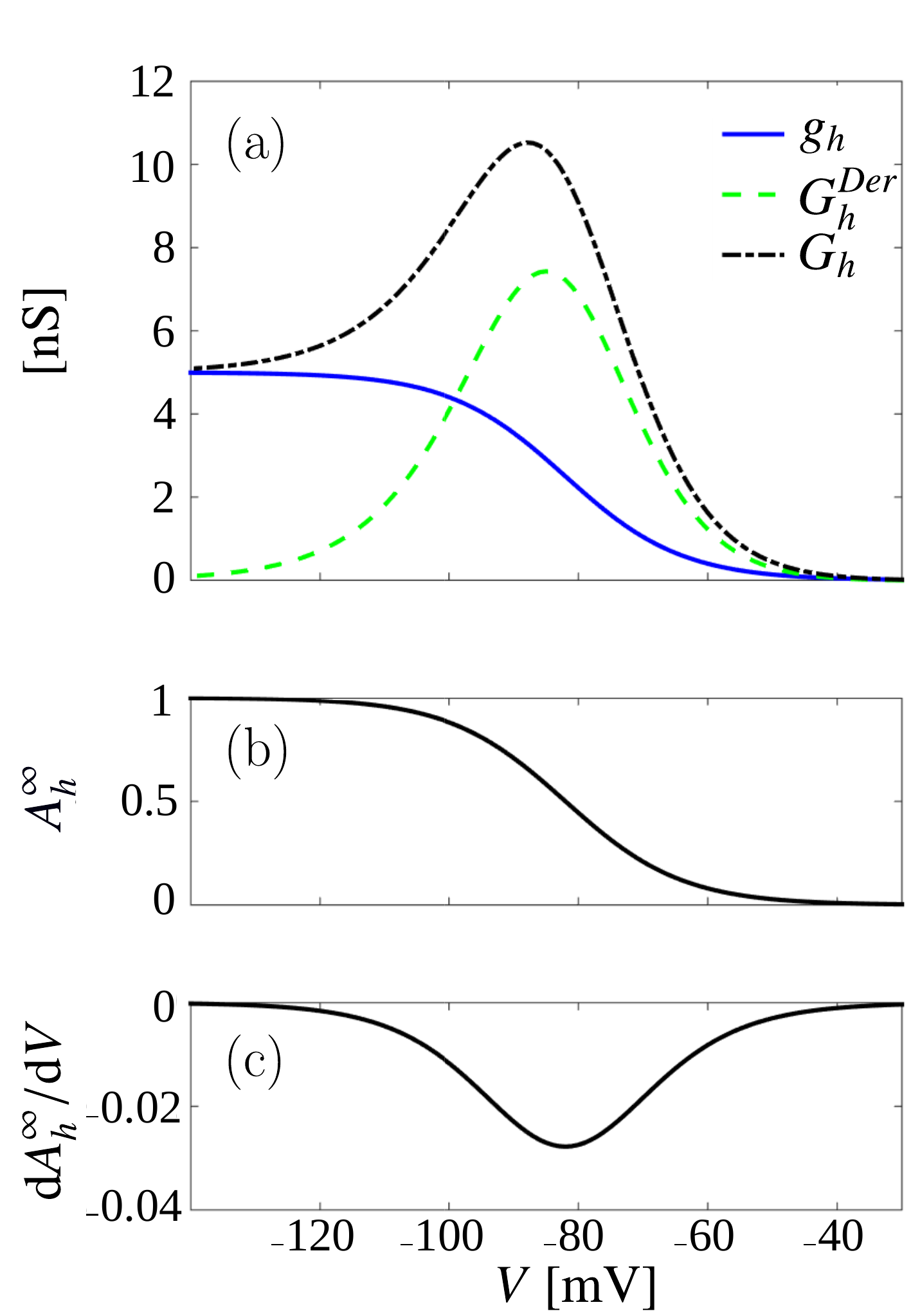}
\caption{\textbf{Slope conductance of $I_h$ and its properties.} (a) Voltage dependence of $I_h$ slope conductance ($G_h$), chord conductance $g_h$ and the derivative conductance $G_h^{Der}$. (b) Voltage dependence of the steady state activation variable $A_h^{\infty}$. (c) Voltage dependence of $dA^{\infty}_{h}/dV$.}
  \label{Fig::fig3}
\end{figure}

Fig.~\ref{Fig::fig3} shows a numerical example of $G_h$, $g_h$, $G_{h}^{Der}$, $A_h^{\infty}$ and $\frac{dA_h^{\infty}}{dV}$ obtained from Eqs.~(\ref{Eq:Ainf}),~(\ref{Eq:slope}) and (\ref{Eq:slope_detailed}). The chord conductance is monotonically decreasing with the membrane potential, since it is directly proportional to the steady state activation variable $A_h^{\infty}$. It approximates asymptotically to the maximum conductance at hyperpolarized voltages and vanishes at depolarized voltages.

Unlike the chord conductance, $G_h^{Der}$ has a non-monotonic behavior and vanishes in two situations: when $A^{\infty}_{h}\rightarrow0$ and $A^{\infty}_{h}\rightarrow1$. Since $G_h^{Der}$ is directly proportional to $\frac{dA_h^{\infty}}{dV}$ and the driving force (see Eq~(\ref{Eq:slope_detailed})), then $G_h^{Der}$ can only have non-vanishing values within the region where the activation changes in a voltage-dependent manner and for membrane potentials far from the reversal potential. This means that $G_h^{Der}$ would contribute significantly for $A^{\infty}_{h}$ values near to 0.5 ($V = V_{1/2} = -82$ mV). In fact, $G_h^{Der}$ has a peak with maximum value near $-82$ mV (Fig.~\ref{Fig::fig3}(a)). $G_h$ reaches its asymptotic value for membrane potentials above $-40$ mV and below $-130$ mV.

$I_h$ has exclusively positive slope conductance for hyperpolarized membrane potentials (see Fig.~\ref{Fig::fig3}(a) and Eqs.~(\ref{Eq:slope}) and (\ref{Eq:slope_detailed})).

\subsection{\label{sec:impedance}Impedance}

The complex impedance is expressed as \cite{kali2012}
\begin{equation}
Z = \frac{1}{g_L + i \omega C + g_h + \frac{G_{h}^{Der}}{1 + i \omega \tau_h}},
\label{Eq::Impedance}
\end{equation}

\noindent where $\omega = 2\pi f$ with $f$ being the stimulation frequency in Hz. We write the impedance magnitude as 
\begin{equation}
|Z| = (Z Z^{*})^{1/2} = \left(A + \omega^2 C^2 + \frac{B - D \omega^2 \tau_h}{1 + \omega^2 \tau_h^2}  \right)^{-1/2} ,
\label{Eq:impedance_magnitude}
\end{equation}

\noindent where $A = (g_L+ g_h)^2$, $B=2G_{h}^{Der}(g_L+ g_h)+(G_{h}^{Der})^2$ and $D=2G_{h}^{Der}C$. Notice that $A$, $B$ and $D$ are positive terms. $A$ depends only on the chord conductance, $B$ depends on both the chord and the derivative conductances, and $D$ depends only on the derivative conductance.

\subsection{\label{sec:phase_plane}Phase plane analysis} 

We used phase plane analysis with variables $A_h$ and $V$ to study how activation of $I_h$ influences voltage responses using the stimulus as explained in section~\ref{sec::simulations}. This has been successfully done elsewhere \cite{rotstein2014b,Mikiel2016}. We plotted $V-A_h$ trajectories that represent one cycle of response to a sinusoidal stimulus at a particular frequency and stimulus amplitude. The $V$-nullcline and $A_h$-nullcline are the curves along which $dV/dt = 0$ and $dA_h/dt = 0$, respectively. For the $V$-nullcline we solved Eq.~(\ref{Eq:model}) and for the gating dynamics we obtained the $A_h$-nullcline from Eq.~(\ref{Eq:Ah_kinetics}).

\subsection{Simulations}
\label{sec::simulations}

Simulations were done in NEURON using the Python interface.
The simulation time step was $0.025$ ms and the initial membrane
potential was $-70$ mV.  The cell specific capacitance is set at $1$ $\mu$F/cm$^{2}$. If not specified, we use $\bar{g}_h = g_L$ = 5 nS. The model has the geometry of a cylinder with $70$ $\mu$m of diameter and $70$ $\mu$m of length, which is chosen to condense the soma and the whole dendritic tree into a single compartment in a manner that preserves the average capacitance of a pyramidal cell ($C\approx 150$ pF) \cite{tamagnini2015}. Moreover, in NEURON all units for conductance are declared in  specific units and internally transformed to non-specific, i.e. [nS/cm$^{2}$] to [nS]. Here we only state the non-specific units but one can simply transform to specific ones by dividing all our units by the area of the cylinder.

Sinusoidal currents were injected using the impedance amplitude profile (ZAP) protocol, which consists of a sinusoidal current with increasing frequency \cite{puil1986}. Here we used a version of this protocol with linearly increasing frequency \cite{cali2008} so that we could evenly study both low and high frequencies,   

\begin{equation}
	I = A\sin[\pi (f(t)-F_\text{start})(t-t_\text{start})],
	\label{eqn::zap_dyn}
\end{equation}

\noindent where $f(t) = F_\text{start} + (F_\text{stop}-F_\text{start})(t-t_\text{start})/(t_\text{stop}-t_\text{start})$, being $F_{\rm start}$ ($F_{\rm stop}$) the initial (final) frequency value of the ZAP current, and $t_{\rm start}$ ($t_{\rm stop}$) the initial (final) time boundaries of the ZAP current.

In order to obtain data from the low stimulation frequencies, we applied a long protocol with duration of $600$ s. The currents had constant amplitude of 10 pA and the frequency $f$ increased linearly in time from $0.001$ to $20$ Hz. To maintain the membrane potential at different values, constant currents were injected.

The resulting voltage response $V(t)$ was measured and the frequency-dependent impedance profile $Z(\omega)$ was calculated as the difference of the absolute value of the amplitude of the voltage peaks and the fixed resting potential. Such measure was normalized by the amplitude of the injected sinusoidal current. The resonance frequency $\omega_{res}$ is defined as the frequency of injected current that maximized $Z(\omega)$. Similarly, we also measure the frequency-dependent activation variable profile $\Delta A_h$ as the difference between the actual activation variable and its initial value. 

\section{\label{sec:results}Results}

\subsection{Existence of resonance and its voltage dependency}

In order to determine analytically the existence properties of resonance, we will obtain a mathematical expression for the frequency at the maximum impedance magnitude. Since resonance occurs when there is a maximum in the impedance magnitude, then by making $\frac{d|Z|}{d\omega}=0= \left( -\frac{|Z|^3}{2} \right) \left( 2\omega C^2 - \frac{2\omega \tau_h(D+B \tau_h)}{(1+\omega^2 \tau_h^2)^2}\right)$ for some $\omega_{res}$ we obtain 
\begin{equation}
\omega_{res}= \frac{\sqrt{\frac{\sqrt{\tau_h \left(D + B\tau_h\right)}}{C}-1}}{\tau_h}.
\label{Eq::resonance_existence}
\end{equation}

Equation (\ref{Eq::resonance_existence}) shows that there is resonance only when $\omega_{res}$ is real, i.e. when $\tau_h \left(D + B\tau_h\right) > C^2$. Otherwise, there is no resonance. Thus, resonance is hampered when $\tau_h$, $B$ or $D$ are small.

The expression for the impedance magnitude, Eq.~(\ref{Eq:impedance_magnitude}), implies that resonance is not present for all voltage values but is restricted to a certain range depending on $\tau_h$. This suggests that both $V$ and $\tau_h$ influence the occurrence of resonance. 

\begin{figure*}[!ht]
\includegraphics[width=1\textwidth]{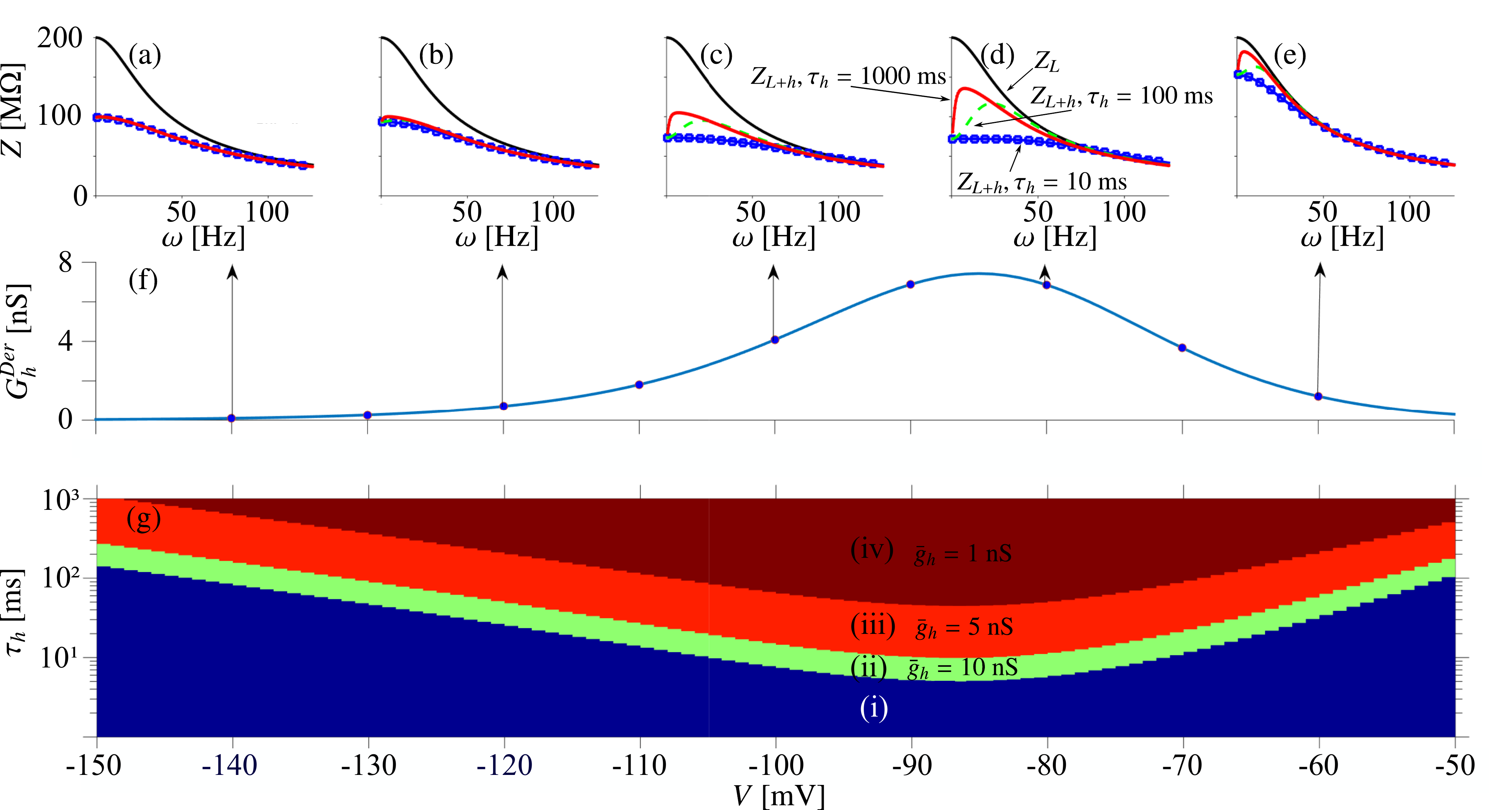}
\caption{\textbf{Impedance dependence of $V$ and $\tau_h$.} [(a)--(e)] Impedance profiles for the passive case (only leak, $Z_L$) and for the case of leak plus $I_h$ ($Z_{L+h}$) for different $\tau_h$ values (10, 100 and 1000 ms) and for different membrane potentials from $V=-60$ to $V=-140$ mV, which is the full range of $I_h$ activation. Notice that resonance peaks are not present for all potentials. (f) $I_h$ derivative conductance ($G_h^{Der}$). The arrows indicate the membrane potential used in the impedance profiles in [(a)--(e)]. (g) Regions where there is resonance. In blue (i) there is no resonance; in (ii) green and above there is resonance for $\overline{g}_h = 10$ nS; in light red (iii) and above, there is resonance for $\overline{g}_h = 5$ nS; and in dark red (iv) there is resonance for $\overline{g}_h = 1$ nS.}
  \label{Fig::fig4}
\end{figure*}

Figure~\ref{Fig::fig4}[(a)--(e)] shows examples of impedance profiles when varying $\tau_h$ and $V$. For $\tau_h = 10$ ms there is no resonance for all membrane potentials considered. For $\tau_h = 100$ ms, there is clear resonance only at $V=-80$ and $V=-100$ mV. For $\tau_h = 1000$ ms resonance is present for all membrane potential, except for $V=-140$ mV.

Figure~\ref{Fig::fig4}(g) shows the parameter regions with resonance in the ($\tau_h$, $V$) diagram when $\overline{g}_h$ is varied. The (i) region in blue corresponds to absence of resonance for all values of $\overline{g}_h$. Notice that below $\tau_h \approx 5$ ms there is no resonance for all membrane potentials. The (ii) region in green and above has resonance when $\overline{g}_h = 10$ nS; the (ii) region in light red and above has resonance when $\overline{g}_h = 5$ nS; and the (iv) region in dark red has resonance when $\overline{g}_h = 1$ nS. Above $\tau_h = 5$ ms it is possible to have resonance, and the bigger $\overline{g}_h$ the larger the area where resonance can occur. Notice also that increasing $\tau_h$ increases the membrane potential range with resonance.

This voltage dependent behavior might be related to the voltage dependence of the chord conductance or the derivative conductance that affects the terms B and D in Eq.~(\ref{Eq::resonance_existence}). The chord conductance increases monotonically with hyperpolarization. However, the resonance behavior is non-monotonic, suggesting that the chord conductance has little influence on it. The derivative conductance has a non-monotonic behavior, as shown in Fig.~\ref{Fig::fig4}(f), which matches the behavior of the resonance. This suggests that the voltage dependent behavior of the resonance is mainly due to the derivative conductance. 

Below we list some evidence that the voltage dependence of the resonance is due mainly to the $G_h^{Der}$ voltage dependence:
\begin{enumerate}
\item In order to influence the impedance profile by means of $\tau_h$, $G_h^{Der}$ must be nonzero. This can be seen from Eq.~(\ref{Eq::Impedance}): if $G_h^{Der} = 0$, the impedance is reduced to $Z = (g_L + i \omega C + g_h)^{-1}$, where $\tau_h$ disappears.

\item Even if $G_h^{Der} = 0$, the impedance remains voltage-dependent due to $g_h$. However, there is no resonance, since when $G_h^{Der} = 0$, $\omega_{res} = \frac{\sqrt{-1}}{\tau_h}$. Phenomenologically, a nonzero $G_h^{Der}$ means that the activation variable $A_h$ is able to vary, since $G_h^{Der}$ is directly proportional to $dA_h^{\infty}/dV$. Thus, the existence of resonance requires a variation of $A_h$ in time.

\item When $g_h = 0$, Eq.~(\ref{Eq::Impedance}) still has a similar form, but with the leak conductance reduced, i.e. setting $g_h = 0$ is equivalent to $g_L'= g_L+g_h$. Then, the system is still able to present resonance. Also, when $g_h = 0$, the term B is reduced to $B=2G_{h}^{Der}g_L+(G_{h}^{Der})^2$, but still has a nonzero value, which means that a resonance frequency $\omega_{res}$ can exist. 
\end{enumerate}

Concluding, in spite of the chord conductance $g_h$ being able to influence resonance, the major responsible for the occurrence of resonance is the derivative conductance $G_h^{Der}$ and not $g_h$.

We also observe that low values of $\tau_h$ or $G_h^{Der}$ abolish resonance. This can be understood as follows: when $\tau_h$ is too small, i.e when $\omega \tau_h \ll 1$, the impedance magnitude is approximately given by $(A + B + \omega^2 C^2)^{-1/2} = ((g_L + g_h + G_{h}^{Der})^2 + \omega^2 C^2)^{-1/2} = ((g_L + G_{h})^2 + \omega^2 C^2)^{-1/2} = (g_L'^2 + \omega^2 C^2)^{-1/2}$. This is equivalent to the impedance of a neuron with only a leak current with a conductance value $g_L'= g_L+G_h$. 

In the case where $G_h^{Der}$ is too small, $B \approx D \approx 0$ in Eq.~(\ref{Eq:impedance_magnitude}) and the impedance magnitude may be approximated by $\left(A + \omega^2 C^2\right)^{-1/2} = ((g_L + g_{h})^2 + \omega^2 C^2)^{-1/2} = (g_L'^2 + \omega^2 C^2)^{-1/2}$. This is equivalent to the impedance of a neuron with only a leak current where the conductance value $g_L'= g_L+g_h$. And when $\tau_h$ or $G_{h}^{Der}$ have low values, the impedance profile is equivalent to the one of an RC circuit, i.e. a low pass filter without resonance.  

\subsection{Phase plane analysis}
\begin{figure*}[!ht]
\includegraphics[width=1\textwidth]{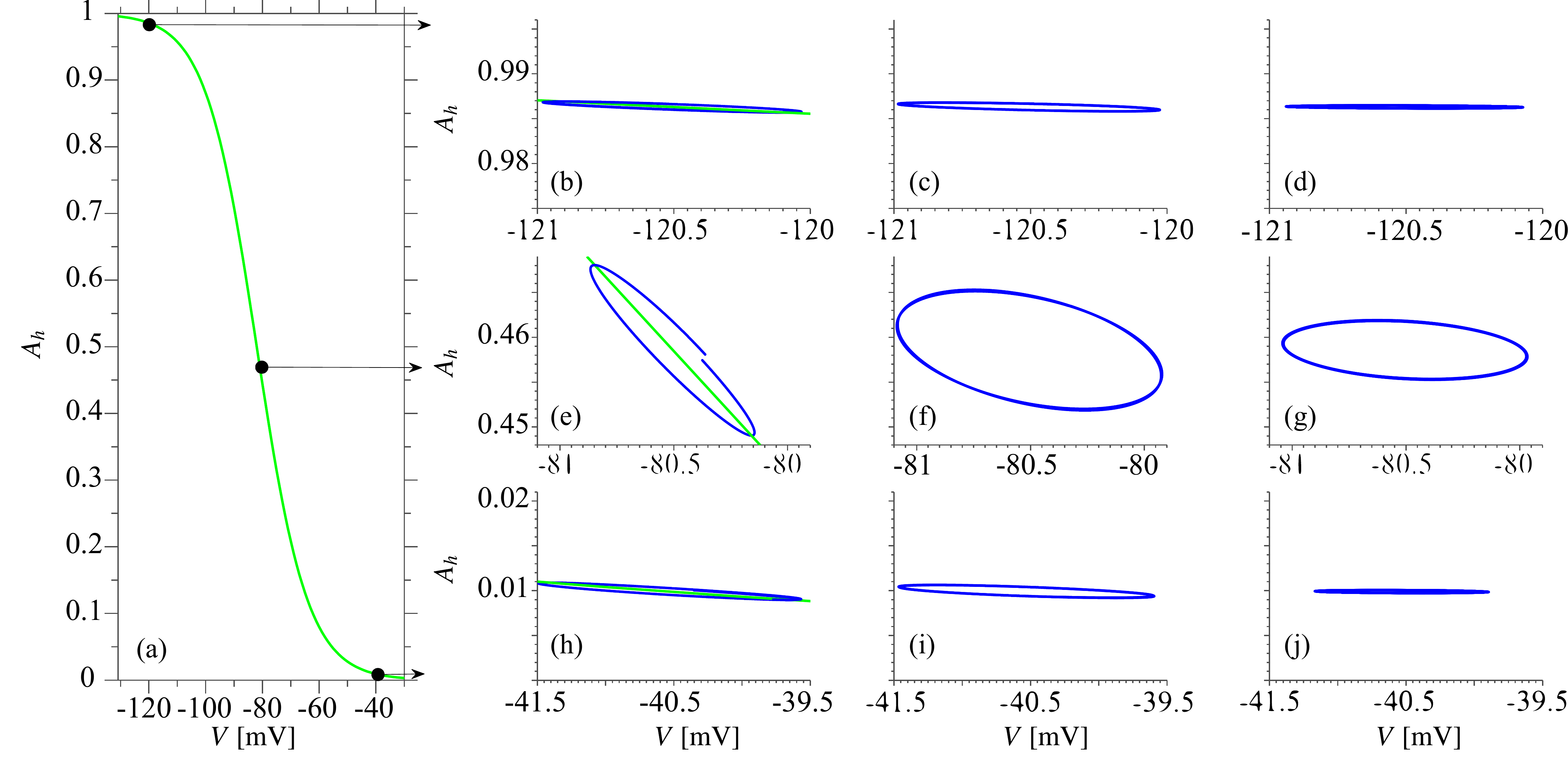}
\caption{\textbf{Evolution of trajectories for different membrane voltages.} (a) Voltage dependence of the steady state activation variable. Horizontal arrows indicate which region is studied on the plots to the right. [(b)--(j)] phase plots of activation variable vs membrane potential of the trajectories for different membrane potentials: [(b)--(d)] $V=-120$ mV, [(e)--(g)] $V=-80$ mV, [(h)--(j)] $V=-40$ mV and for different stimulation frequencies ($\omega$): [(b),(e), and (h)] low frequency, [(e),(f), and (i)] resonance frequency $\omega_{res}$, and [(d),(g), and (j)] high frequency. The green (light gray) line represents the slope of $dA^{\infty}_h/dV \propto G^{Der}_h$. In all panels $\tau_h=50$ ms}
  \label{Fig::figisoclinas}
\end{figure*}

\begin{figure*}[!ht]
\includegraphics[width=1\textwidth]{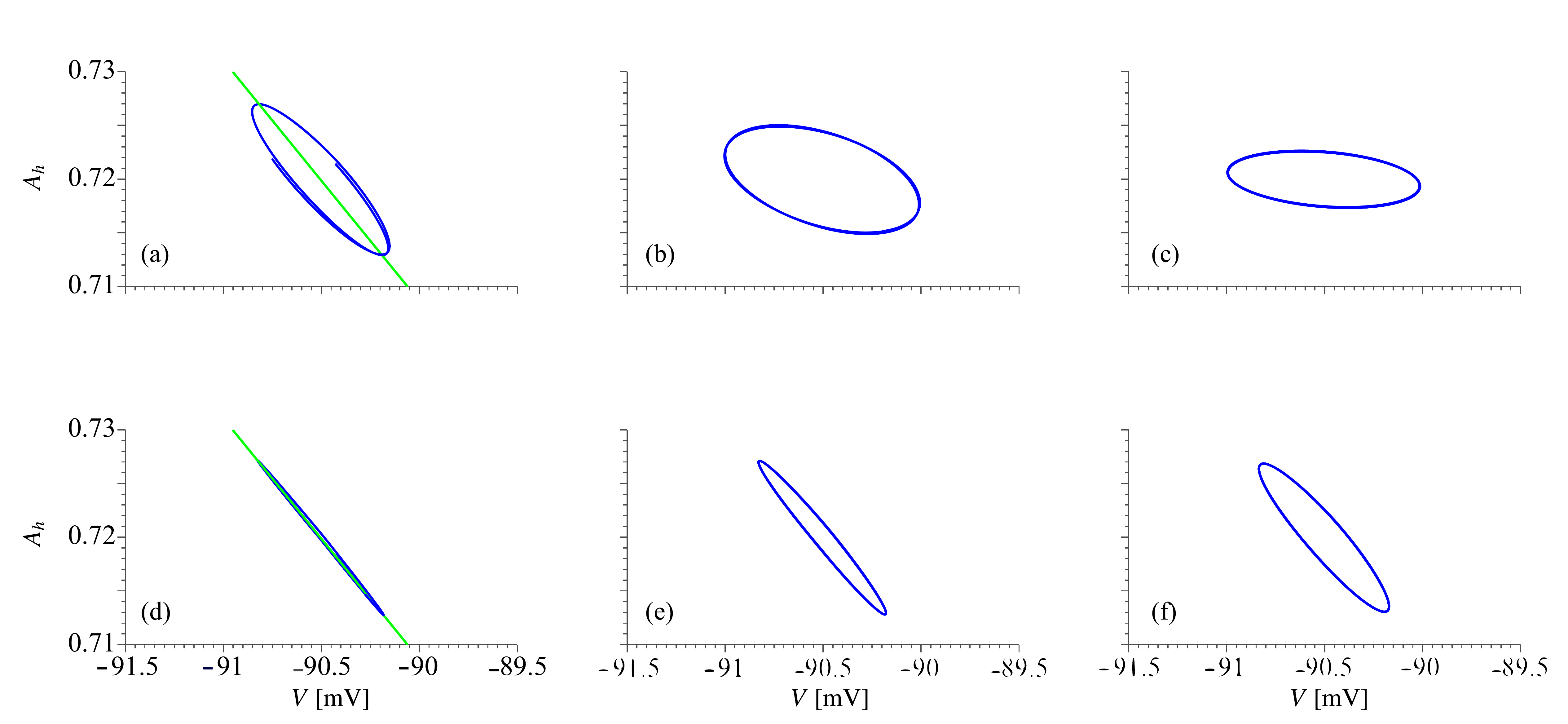}
\caption{\textbf{Evolution of trajectories for different $\tau_h$.} Phase plots of activation variable versus membrane potential of the trajectories for different activation time constant: [(a)--(c)]  $\tau_h=50$, and [(d)--(f)] $\tau_h=5$ ms. There are different stimulation frequencies ($\omega$): [(a) and (d)] low frequency, [(b) and (e)] resonance frequency $\omega_{res}$, and [(c) and (f)] high frequency. The membrane potential was fixed at $V=-90$ mV. Green (light gray) line slope represents the $dA^{\infty}_h/dV \propto G^{Der}_h$.}
  \label{Fig::figisoclinas_tau}
\end{figure*}

The phase plane analysis in the $A_h-V$ diagram is shown in Figs.~\ref{Fig::figisoclinas} and~\ref{Fig::figisoclinas_tau}. In the case of Fig.~\ref{Fig::figisoclinas}, we plot the trajectories fixing the membrane potential with an external constant current so that the steady voltage is $V=-120,-80$, and $-40 \rm mV$ (indicated by horizontal arrows in the figure). In Fig.~\ref{Fig::figisoclinas}(a) we plot $A_h(V)$, which stays in three different regions of the curve depending on the voltage. These regions  correspond to the cases where $I_h$ is fully activated, half-activated and almost deactivated. Only the case with $V=-80 \rm mV$ displays resonance. In addition, we plot the trajectories for three different frequencies: ($\omega =\omega_{res}$), a lower frequency ($\omega < \omega_{res}$), and a higher frequency ($\omega > \omega_{res}$). In Fig.~\ref{Fig::figisoclinas} we fixed the value of $\tau_h = 50$ ms. For low frequencies, regardless of the voltage value, the trajectory follows the $A_h$-nullcine, which is depicted in green (light gray) in the plots of low-frequencies (Fig.~\ref{Fig::figisoclinas}[(b), (e), and (h)]). This means that $I_h$ is able to track the slow changes of the voltage, being fully activated and deactivated during the full cycle. 

In contrast, for high frequencies, again regardless of the voltage value, the trajectory is mainly horizontal, which means that $I_h$ is not able to track the fast changes of the voltage. Then the activation variable $A_h$ changes slightly. 

In regard to the voltage, when it is too depolarized or hyperpolarized ($V=-40$ mV and $V=-120$ mV, respectively), the trajectories are mainly horizontal, regardless of the frequency value. In contrast, for $V=-80$ mV, $G_h^{Der}$ is close to its maximum, the trajectory starts oblique and thin, but by increasing the frequency it gets more horizontal and round.  

In Fig.~\ref{Fig::figisoclinas_tau} we fixed the voltage at $-90 \rm mV$ and vary $\tau_h$ to $5$ and $50$ ms. Only for the case of $\tau_h$ = 50 ms (Fig.~\ref{Fig::figisoclinas_tau}[(a)--(c)]) there is resonance. One can observe the same tendencies as in Fig.~\ref{Fig::figisoclinas}. Interestingly, for the case with $\tau_h = 5$ ms, the trajectory is kept oblique for the high frequency, which means that $I_h$ is still able to track the fast changes of the voltage (Fig.~\ref{Fig::figisoclinas_tau}[(d)--(f)]). 

Notice that for high $\tau_h$, the trajectory starts oblique and thin, but increasing the frequency makes the trajectory more horizontal and round, and this occurs more rapidly than in the case of low $\tau_h$.

In summary, there are two different scenarios where the existence of resonance is hampered. First, when the trajectory starts horizontal for low $\omega$ and does not change regardless of the $\omega$ value, and, second, when the trajectory starts oblique for low $\omega$ and also does not change regardless of the evolution of $\omega$. The former situation is caused by low $G^{Der}_h$ values, i.e., low $dA^{\infty}_h/dV$ values. The latter situation is caused by low $\tau_h$ values that keep $dA_h/dV$ invariant to changes in $\omega$. This allows us to conclude that resonance emerges for trajectories that start oblique (i.e., with high $dA^{\infty}_h/dV$ values) and thin but rapidly become horizontal and round (due to high $\tau_h$ values) with the variation of $\omega$.

\subsection{Resonance frequency and its voltage dependency}

\begin{figure*}[!ht]
\includegraphics[width=1\textwidth]{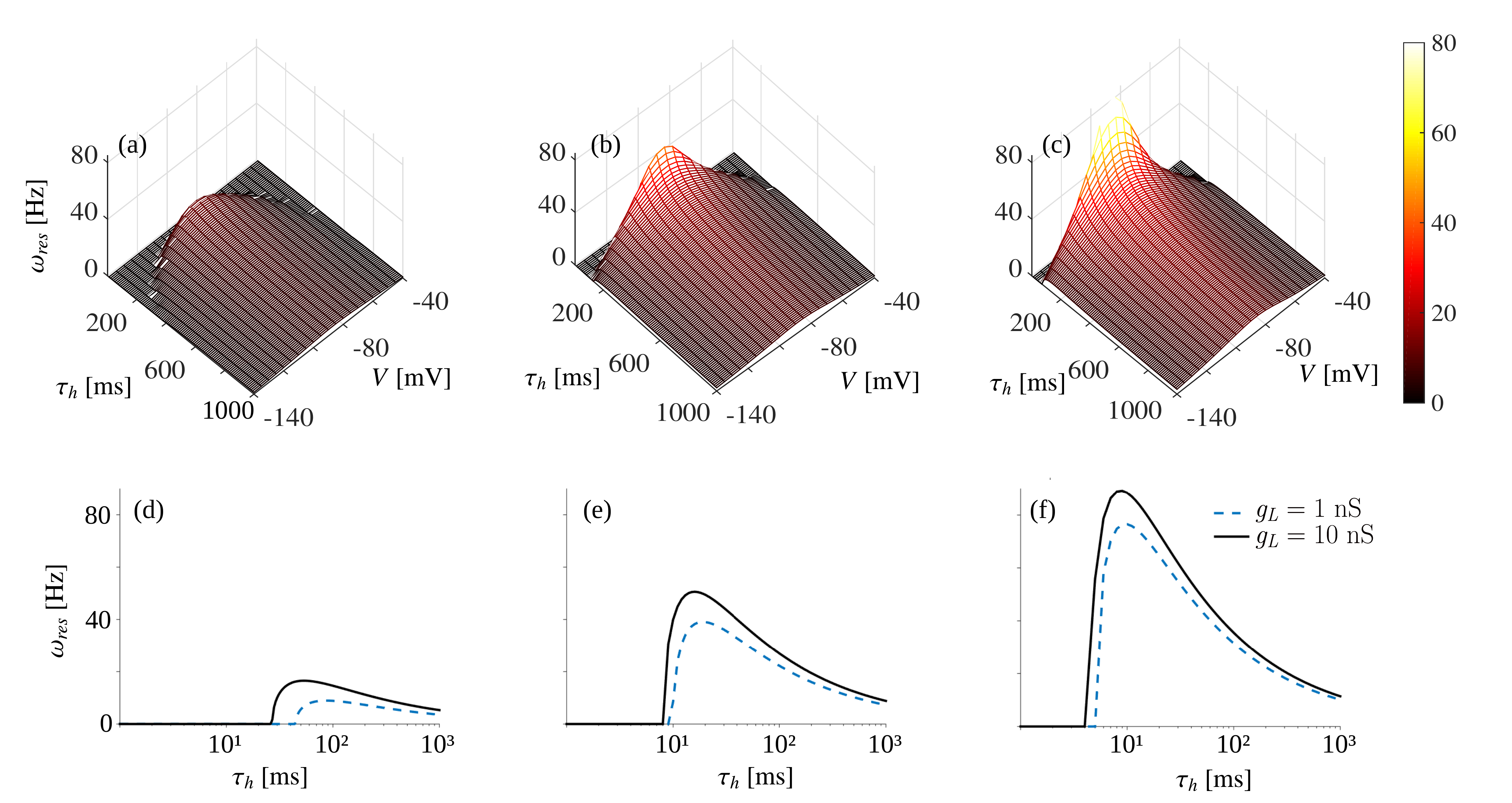}
\caption{\textbf{Resonance frequency dependence with $\tau_h$ and $V$.} [(a)--(c)] Resonance frequency ($\omega_{res}$) for different values of $\tau_h$ and $V$. [(d)--(f)] Resonance frequency ($\omega_{res}$) for different values of $\tau_h$ and $g_L$. [(a) and (d)] $\bar{g}_h=1$ nS, [(b) and (e)] $\bar{g}_h=5$ nS, and [(c) and (f)] $\bar{g}_h=10$ nS. In all cases $V = -85$ mV.}
\label{Fig::fig5}
\end{figure*}

In the previous sections we determined the main biophysical parameters of $I_h$ underlying the existence of resonance. In this section we will determine how the parameters $\overline{g}_h$, $g_L$, $\tau_h$ and V affect the value of the resonance frequency ($\omega_{res}$) using Eq.~(\ref{Eq::resonance_existence}). Clearly, increasing $B$ or $D$ increases the value of $\omega_{res}$ while increasing $\tau_h$ decreases the value of $\omega_{res}$. Since B increases when $G_h^{Der}$, $g_h$ or $g_L$ increase, and since D increases when $G_h^{Der}$ increases, we conclude that the increase of any type of conductance (i.e. leak, $I_h$ chord or $I_h$ derivative conductance) increases the value of the resonance frequency. However, the contribution of $g_h$ or $g_L$ to resonance is modulated by $G_h^{Der}$ (see the first term in $B=2G_{h}^{Der}(g_L+ g_h)+(G_{h}^{Der})^2$). This is reflected in the numerical values as shown in Fig.~\ref{Fig::fig5}[(a)--(c)]. Fig.~\ref{Fig::fig5} shows the resonance frequency for different values of $\tau_h$ and $V$, when $\overline{g}_h$ is varied. Notice that the resonance frequency has a voltage dependency with a maximum value near $-90 \rm mV$ and low values for depolarized and hyperpolarized membrane potentials, resembling the voltage-dependent behavior of $G_h^{Der}$. This trend strongly suggests that $G_h^{Der}$ is the main factor determining the voltage dependency of $\omega_{res}$.   

In Fig.~\ref{Fig::fig5}[(d)--(f)] we show the value of the resonance frequency when the membrane potential is fixed at $V = -85$ mV, which is a membrane potential close to the highest resonance frequency (see  Fig.~\ref{Fig::fig5}[(a)--(c)]).
Notice that changing the leak conductance from $1$ nS to $10$ nS increases the resonance frequency ($\approx$ 10 Hz), but the change is smaller than that observed when the $I_h$ conductance is changed in the same proportion.  

But what is the mechanism by which $I_h$ determines the resonance frequency? Based on the results presented above, we propose that it is the magnitude of the change of the activation variable in time $dA_h/dt$ that determines the resonance frequency. This would explain why increasing $G_h^{Der}$ increases the resonance frequency while increasing $\tau_h$ decreases the frequency resonance, as suggested by Eq.~(\ref{Eq::resonance_existence}).

\begin{figure}[!ht]
\includegraphics[width=0.45\textwidth]{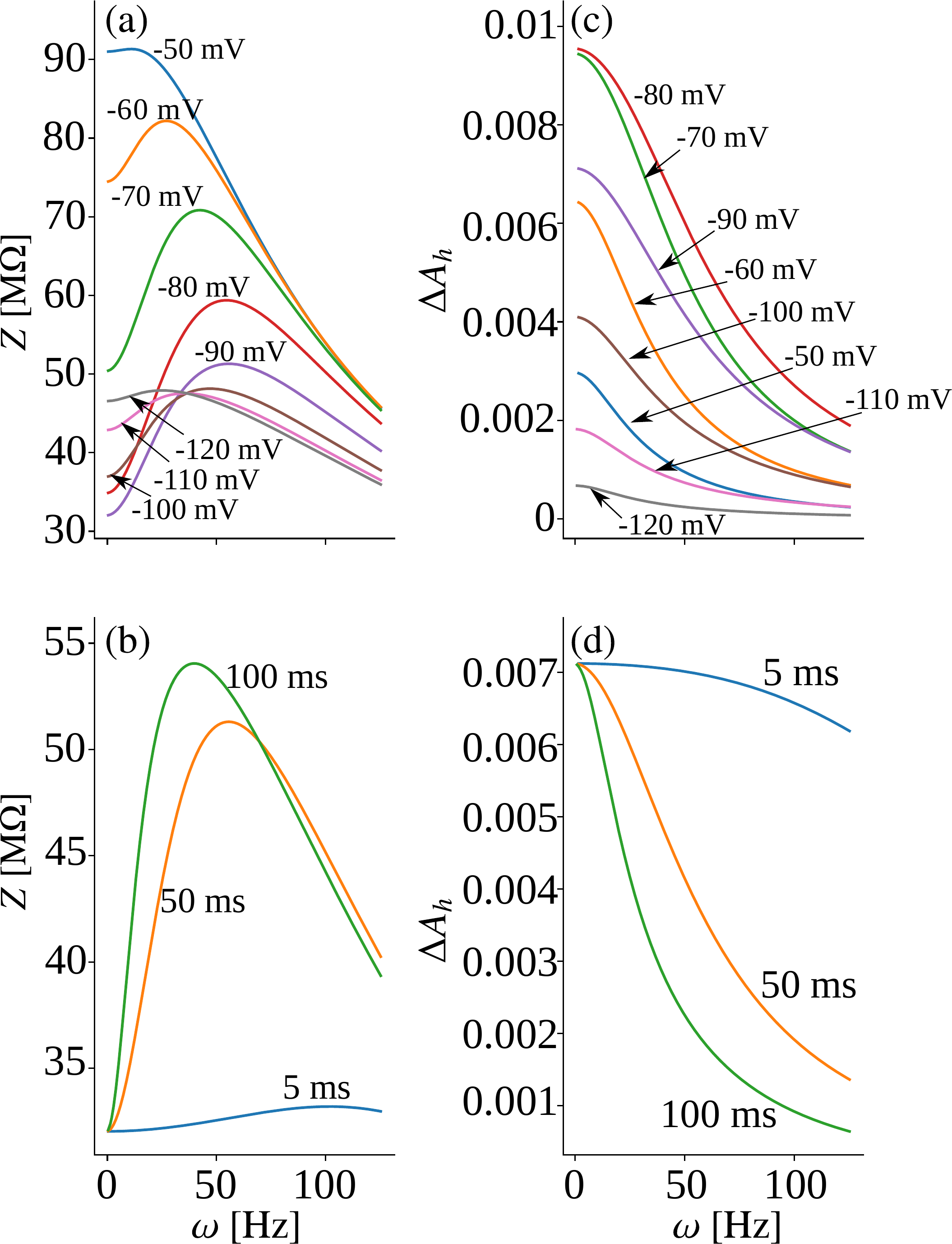}
\caption{\textbf{Impedance profiles and their dependency with $V$ and $\tau_h$.} (a) Impedance profiles for different membrane potentials, and (b) for different $\tau_h$. [(c)--(d)] Variation of $A_h$. In [(a) and (c)] $\tau_h = 50$ ms.  In [(b) and (d)] we fixed  $V = -90 \rm mV$ and varied $\tau_h$ (5, 50, and 100 ms).}
  \label{Fig::fig6}
\end{figure}

In order to verify this hypothesis, we analyzed the impedance profiles obtained from computational simulations where ZAP currents where injected into a single compartment neuron model with a leak current and $I_h$. Figure~\ref{Fig::fig6}(c) shows the variation of the activation variable for different membrane potentials, sweeping the $I_h$ activation range. Note that the maximum variation of $A_h$ is observed for $V=-80$ mV (see Figure~\ref{Fig::fig6}(a)), i.e. the membrane potential for the maximum $G_h^{Der}$. Moreover, $\Delta A_h$ decreases monotonically with $\omega$. Fig.~\ref{Fig::fig6}(d) shows the variation of the activation variable for different $\tau_h$ values. Notice that the initial $\Delta A_h$ is the same for all $\tau_h$ values but each curve evolves differently, where the curve with the lower $\tau_h$ ($5$ ms) has a slower decay with the frequency. 

It is well known that the resonance frequency is related to the frequency-dependent attenuation due to $I_h$ activation. $I_h$ acts as a high pass filter, attenuating strongly slow voltage changes but not affecting fast voltage changes. Our results suggest that $G_h^{Der}$ and $\tau_h$ are the main factors that influence this attenuation. Our hypothesis is that $G_h^{Der}$ and $\tau_h$ determine the variation of the activation variable in time ($dA_h/dt$), and such variation determines the magnitude of the attenuation due to $I_h$. If this is true, then the increase of $dA_h/dt$ should increase the attenuation of the voltage changes by $I_h$ and, consequently, increase the resonance frequency.

\begin{figure}[!ht]
\includegraphics[width=0.5\textwidth]{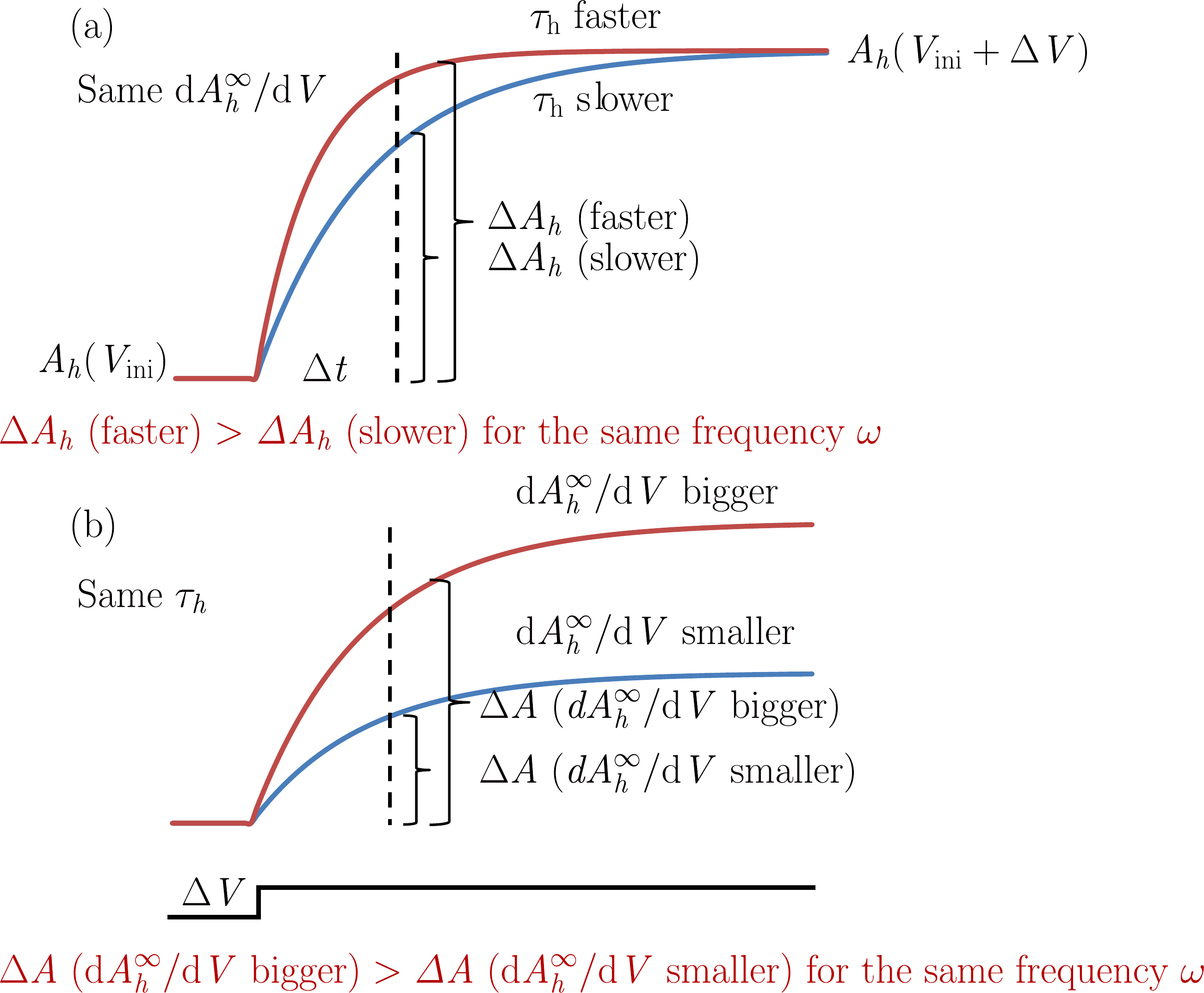}
\caption{Schematic diagram explaining the mechanism by which the interaction between frequency and activation variable time evolution determines the amount of the change of the activation variable $\Delta A_h$. The maximum amplitude of the change of the activation variable is determined by $dA_h^{\infty}/dV$, while the time evolution is determined by $\tau_h$. (a) For the same value of $dA_h^{\infty}/dV$, $\Delta A_h$ is bigger for the faster $\tau_h$ values. (b) For the same $\tau_h$, $\Delta A_h$ is bigger for the bigger $dA_h^{\infty}/dV$ values.}
\label{Fig::fig7}
\end{figure}

But how $\tau_h$ determines the resonance frequency? The change of $A_h$ in time is not instantaneous but obeys some dynamics which is determined by $\tau_h$ (Fig.~\ref{Fig::fig7}(a)). The slower the $\tau_h$ the smaller the change of $A_h$ in time, i.e. $dA_h/dt$ is smaller for bigger $\tau_h$. Increasing $\tau_h$ implies a decrease in $dA_h/dt$. Such decrease leads to a decrease of the $I_h$ attenuation, thus decreasing the resonance frequency. In agreement with this, notice that decreasing $\tau_h$ decreases the impedance magnitude (Fig.~\ref{Fig::fig6}(b)), i.e., decreasing $\tau_h$ enhances the $I_h$ attenuation. 

Figure~\ref{Fig::fig6}(d) shows that $A_h$ with slower $\tau_h$ decays faster than $A_h$ with faster $\tau_h$, and this tendency is reflected by $\omega_{res}$, where $A_h$ with faster decay has an impedance profile with lower $\omega_{res}$. 

But how $G_h^{Der}$ determines the resonance frequency? $G_h^{Der}$ is directly proportional to $dA_h^{\infty}/dV$, i.e. the maximum variation of the activation variable for a change of the voltage. Since the magnitude of $dA_h/dt$ is directly proportional to $dA_h^{\infty}/dV$, then increasing $G_h^{Der}$ implies an increase in $dA_h/dt$ due to an increase in $dA_h^{\infty}/dV$ (Fig.~\ref{Fig::fig7}(b)). This increase in $dA_h/dt$ leads to an increase of the $I_h$ attenuation, thus increasing the resonance frequency. Accordingly, this tendency can be observed in Fig.~\ref{Fig::fig6}(a) at $\omega = 0$. Notice that the impedance profiles with higher attenuation correspond to higher $\Delta A_h$ values. 

More specifically, when $\omega \rightarrow 0$, the impedance defined in Eq.~(\ref{Eq:impedance_magnitude}) becomes $|Z| = \left(A + B\right)^{-1/2}$. By rearranging the terms we have that $\lim\limits_{\omega \to 0} |Z| = (g_L + G_{h})^{-1}$. This result demonstrates that the impedance magnitude when $\omega \rightarrow 0$ is independent of $\tau_{h}$, but voltage-dependent following the $I_h$ slope conductance ($G_h$). In the same equation, one also see that the increase of $g_L$ or $\bar{g}_h$ decreases the impedance magnitude when $\omega \rightarrow 0$. 

Consistent with our hypotheses, variations of $\tau_h$ change the variation of $A_h$ for high frequencies (see Fig.~\ref{Fig::fig6}(d)) in a manner in which increasing $\tau_h$ decreases the variation of $A_h$.

Summarizing, we conclude that the main factors that determine the resonance frequency are $G_h^{Der}$ and $\tau_h$ acting on the change of the activation variable in time ($dA_h/dt$).

\begin{figure}[!ht]
\includegraphics[width=0.5\textwidth]{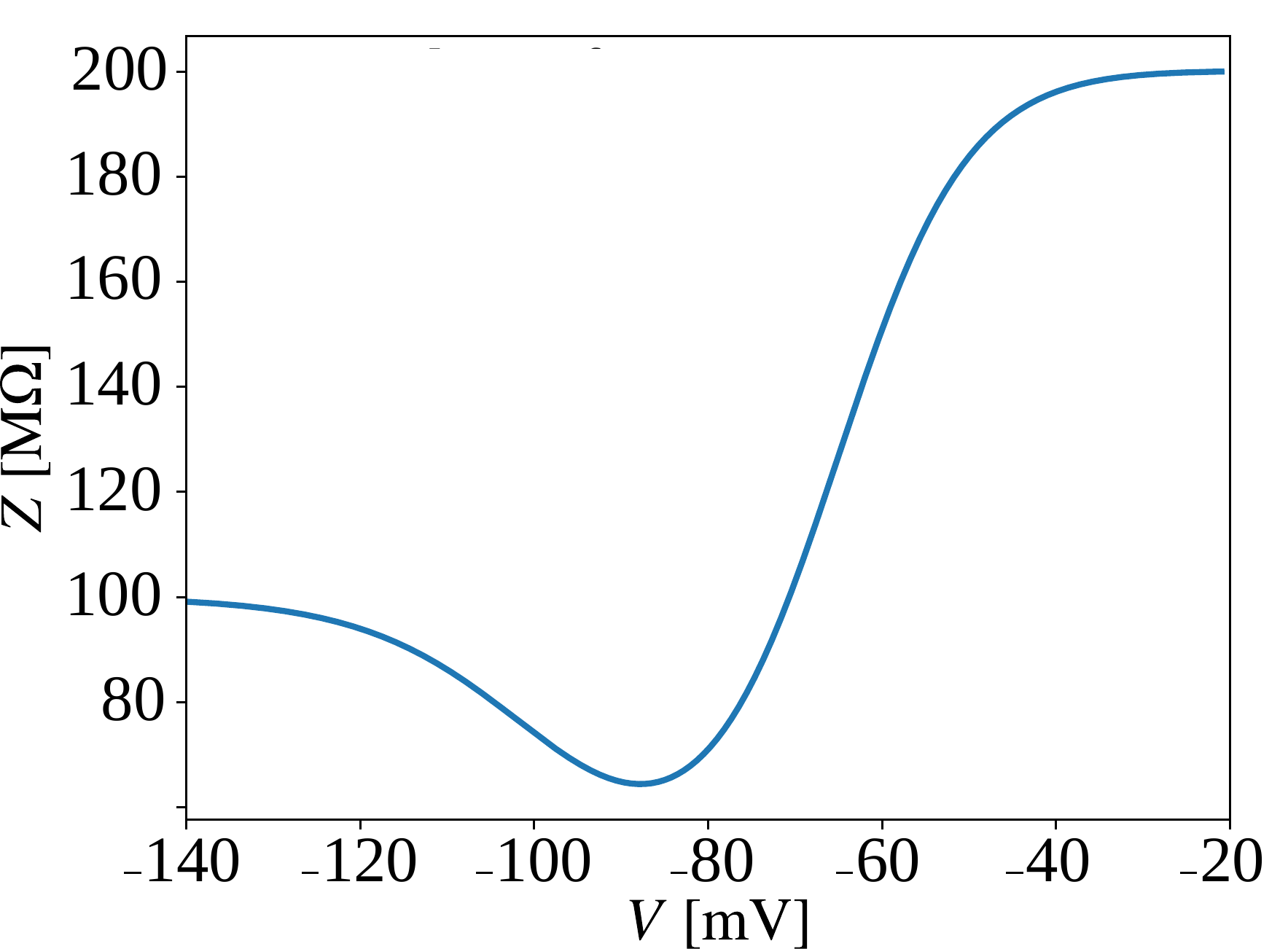}
\caption{\textbf{Impedance at vanishing frequency is voltage-dependent.} Impedance magnitude when $\omega = 0$ Hz.}
  \label{Fig::figX}
\end{figure}

\subsection{Impedance attenuation by $I_h$ at low frequencies}

It is well known that $I_h$ behaves as a high pass filter, attenuating slow changes of the voltage. Figs.~\ref{Fig::fig4}[(a)--(e)] and~\ref{Fig::fig6} show that the variation of the impedance profile with $I_h$ displays a band-pass behavior due to an attenuation of the impedance magnitude at low frequencies, and this attenuation is also influenced by $\tau_h$. This is clearly observed in Fig.~\ref{Fig::fig4}[(a)--(e)] when comparing the impedance magnitude of a neuron model with only leak ($Z_L$) current with a neuron model with leak plus $I_h$ ($Z_{L+h}$) current. Accordingly, one can observe in all $Z_{L+h}$ curves the same attenuation at low frequencies. At first, this attenuation seems to cover the whole range of frequencies. However, when we zoom in we observe that depending on the $\tau_h$ value some impedance curves for the $Z_{L+h}$ case are unexpectedly amplified at high frequencies, when compared with the $Z_L$ case. Thus, in order to determine the behavior of the $I_h$ attenuation on the impedance magnitude (see 
Eq.~(\ref{Eq:impedance_magnitude})) for different stimulation frequencies, we will compare the impedance profiles of the passive case (i.e., only leak current) and the case with $I_h$. Our interest is to determine the regions where $I_h$ attenuates the impedance magnitude and the regions where $I_h$ amplifies it. Thus, by making $|Z_L| = |Z_{L+h}|$ we will be able to study the existence of crossing frequencies ($\omega_c$), and describe the regions with attenuation by comparing the curves. 

\begin{figure}[!ht]
\includegraphics[width=0.5\textwidth]{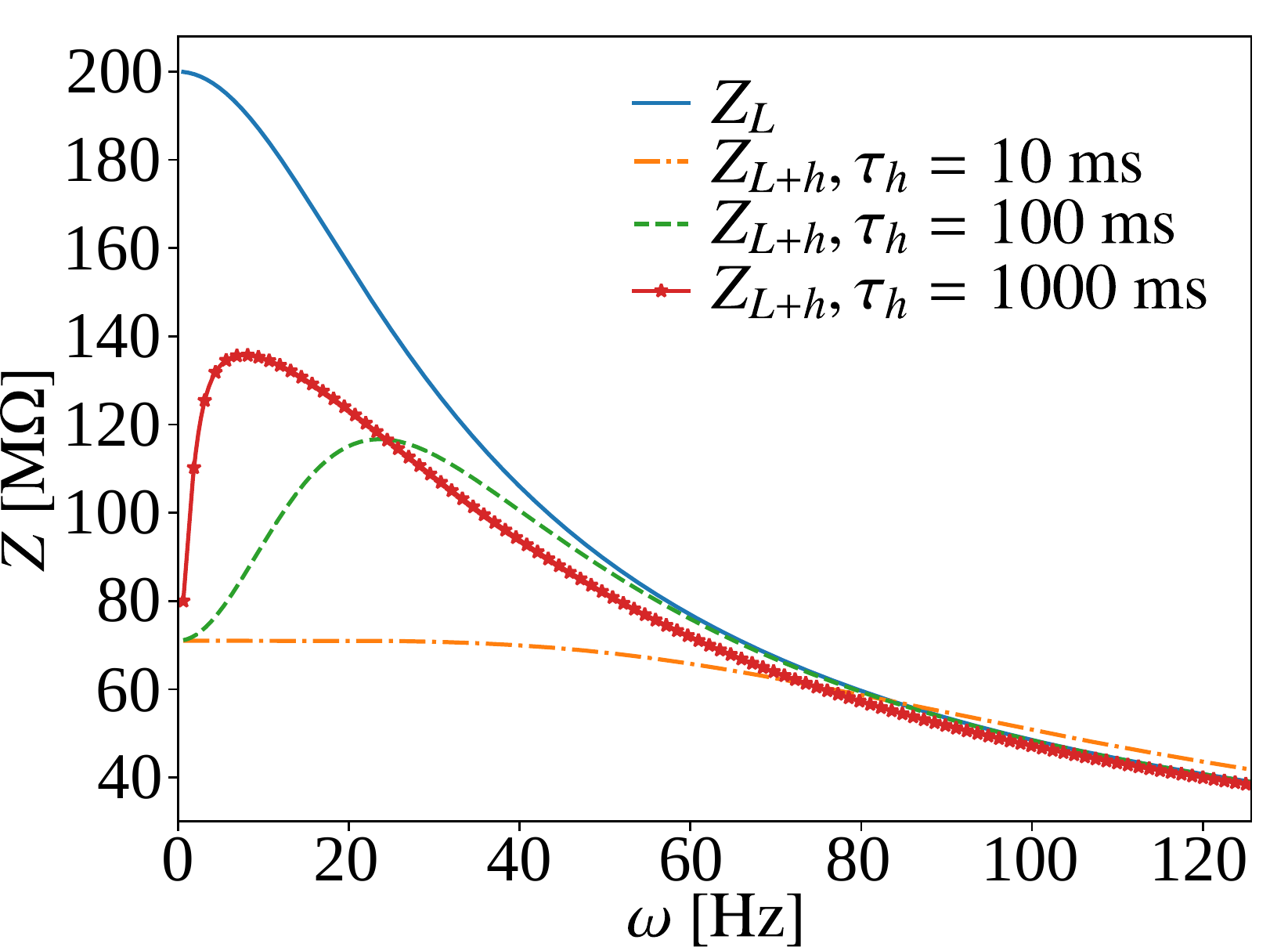}
\caption{\textbf{Impedance profiles.} Impedance magnitude for $V = -80$ mV. The blue (solid) curve is the case with only a leak current. Orange(dotted and dashed), green (dashed) and red (solid with stars) curves correspond to the cases with leak current plus $I_h$ with $\tau_h = 10, 100$ and $1000$ ms, respectively. The green (dashed) and red (solid with stars) curves cross at $\approx 25$ Hz, the orange (dotted and dashed) and blue (solid) curves cross at $\approx 85$ Hz, and the red (solid with stars) and blue (solid) curves never cross.}
  \label{Fig::fig2}
\end{figure}

In this regard, we obtain

\begin{equation}
 \left(g_L ^2 + \omega^2 C^2\right)^{-1/2} = \left(A + \omega^2 C^2 + \frac{B - D \omega^2 \tau_h}{1 + \omega^2 \tau_h^2} \right)^{-1/2},
 \label{Eq:Zl_equal_Zlh}
\end{equation}

\noindent and with a simple rearrangement of terms we isolate the crossing frequency $\omega_c$:
\begin{equation}
\omega_c = \sqrt{\frac{B+E}{D \tau_{h} - E \tau_{h}^2}}, 
\label{Eq:cross}
\end{equation}

\noindent where $E = 2g_{L} g_{h} + g_{h} ^2$. Note that the solution in 
Eq.~(\ref{Eq:cross}) has some constraints due to a subtraction in the denominator as well as the existence of a square root. If the solution is real (i.e., when $D > E\tau_h$) the curves cross once at some frequency ($\omega_c$), otherwise the curves do not cross. Examples of both cases are shown in Fig.~\ref{Fig::fig2}. As a general trend, one observes that the profiles $|Z_L|$ and $|Z_{L+h}|$ can cross or not, but two profiles $|Z_{L+h}|$ with different $\tau_h$ always cross each other (see below).    

Since there cannot be more than one crossing point, one curve will be above the other on one side of $\omega_c$ and the curves switch position on the other side of $\omega_c$. Then, if one evaluates any values of the curves on one side of $\omega_c$ and checks their relative positions, one can infer all the other relative positions with respect to $\omega_c$. 

Using this approach, we will demonstrate that $I_h$ always attenuates the impedance magnitude at low frequencies but not always for high frequencies. We know that for $\omega < \sqrt{\frac{B}{D\tau_{h}}}$, both the left-hand and the right-hand sides of 
Eq.~(\ref{Eq:Zl_equal_Zlh}) are positive and the sum of the terms inside the square root is higher on the right-hand side than on the left-hand side. As a consequence of this, the impedance magnitude of the right-hand side (i.e. with leak current plus $I_h$) is lower than the impedance magnitude of the left-hand side (i.e. with only leak current) for low frequencies (Fig. \ref{Fig::fig2}). The same conclusion could be achieved if one studied the limit $\omega \to 0$ as one can see in Fig.~\ref{Fig::figX}.

The addition of an $I_h$ current decreases the impedance magnitude for low frequencies when compared with the case with only leak current. This implies that for $\omega > \omega_c$, $I_h$ increases the impedance magnitude (Fig. \ref{Fig::fig2}). Moreover, if there is no crossing frequency, then adding $I_h$ decreases the impedance magnitude for all frequencies.

Figures~\ref{Fig::fig4} and~\ref{Fig::fig2} show that two profiles $|Z_{L+h}|$ with different $\tau_h$ always cross each other. Thus, one can determine the influence of $\tau_h$ on the $I_h$ attenuation of the impedance magnitude (Eq.~(\ref{Eq:impedance_magnitude})). To do this, one can compare two impedance profiles having different $\tau_h$ values: $\tau_{h,1}$ and $\tau_{h,2}$, for example. Following the same reasoning as above for Eq.~(\ref{Eq:Zl_equal_Zlh}), we take Eq.~(\ref{Eq:impedance_magnitude}) and look for crossing points, i.e. points which obey $|Z(\tau_{h,1})|= |Z(\tau_{h,2})|$. This gives,
\begin{equation}
\frac{B - D \omega^2 \tau_{h,1}}{1 + \omega^2 \tau_{h,1}^2}  = \frac{B - D \omega^2 \tau_{h,2}}{1 + \omega^2 \tau_{h,2}^2},
 \label{Eq:Zhtau1_equal_Zhtau2}
\end{equation}

\noindent and the crossing frequency ($\omega_c$) can be determined as:
\begin{equation}
\omega_c = \sqrt{\frac{B(\tau_{h,1}+\tau_{h,2})+D}{D \tau_{h,1} \tau_{h,2}}}. 
\label{Eq:two_taus_crossFreq}
\end{equation}

The term inside the square root in Eq.~(\ref{Eq:two_taus_crossFreq}) is always positive, implying the solution is always real and there is always a crossing point (only one) when two impedance profiles with different values of $\tau_h$ are compared.

Additionally, we can check which curve is on top of each other before and after $\omega_c$ is crossed, as before. If $\tau_{h,1}>\tau_{h,2}$ and $\omega < \sqrt{\frac{B}{D\tau_{h,1}}}$, both the left and the right-hand sides of 
Eq.~(\ref{Eq:Zhtau1_equal_Zhtau2}) are positive and the numerator of the term on the left-hand side is lower than the numerator of the term on the right-hand side. Also, the denominator of the term on the left-hand side is higher than the denominator of the term on the right-hand side. Thus, the value on the left-hand side is smaller than the value on the right-hand. As a result, substituting the terms in Eq.~(\ref{Eq:Zhtau1_equal_Zhtau2}) into 
Eq.~(\ref{Eq:impedance_magnitude}), the impedance magnitude of the left-hand side (i.e. with $\tau_{h,1}$) is higher than the impedance magnitude of the right-hand side (i.e. with $\tau_{h,2}$) for low frequencies (Fig.~\ref{Fig::fig2}).

Therefore, we conclude that decreasing $\tau_h$ decreases the impedance magnitude for low frequencies. On the other hand, this implies that for $\omega > \omega_c$ (at high frequencies) the impedance behavior is inverted implying that a decrease in $\tau_h$ will increase the impedance magnitude. These observations are shown in the examples of Fig.~\ref{Fig::fig2}.

\subsection{Effect of instantaneous currents on resonance properties}
The model used with leakage and $I_h$ currents is useful to understand the resonance properties of the neuron. However, real neurons have other voltage-dependent subthreshold currents besides the resonator currents. A special case which has been well characterized is the one when these currents have almost instantaneous activation \cite{hutcheon1996a,hutcheon1996b,hutcheon2000}. The persistent sodium current ($I_{NaP}$) and the inwardly rectifying potassium current ($I_{KIR}$) are two types of currents with these features that are expressed in cortical and other types of neurons \cite{ceballos2017,hutcheon1996a,hutcheon1996b}.

When $\tau = 0$, the impedance can be written as $Z = (g_L + G_i + i \omega C + g_h + G_{h}^{Der}/(1 + i \omega \tau_h))^{-1}$, where $G_i$ is the slope conductance of the instantaneous current. Then, if we rewrite this equation using $g_L' = g_L + G_i$ we notice that the main effect of an instantaneous current on the impedance is an increase or decrease of the leak conductance depending on the sign of $G_i$. For instance, in the subthreshold range $G_{NaP}$ is negative \cite{ceballos2017} while $G_{KIR}$ can be positive or negative. If $G_i$ is positive $g_L'$ will increase, and if $G_i$ is negative $g_L'$ will decrease. The predictions on changes on leak conductances is shown in Fig.~\ref{Fig::fig5}.    

\section{\label{sec:discussion}Discussion}
The objective of this paper was to determine the role of the derivative conductance and the current kinetics on the resonance properties of neurons with leakage and $I_h$ currents. Our main finding is that the interplay between the derivative conductance and the current kinetics, represented by $dA_h/dt$, determines the existence of the resonance and the resonance frequency. 

We found that increasing $G_h^{Der}$ or $\tau_h$ increases the likelihood of occurrence of resonance. Resonance can arise even at a low value of each one of these parameters if the value of the other is high. For instance, neurons from the auditory brainstem display resonance created by a fast activated potassium current ($\tau_K <$ 1 ms) with a high conductance value ($\overline{g}_K = 190$ nS) \cite{Mikiel2016}. In contrast, CA1 pyramidal cells display resonance created by the slow $I_h$ ($\tau_h >$ 10 ms up to 1 s) with a low conductance value ($\overline{g}_h = 5$ nS). 

According to Hutcheon and Yarom \cite{hutcheon2000}, resonance emerges when a current with a positive slope conductance has an activation time constant bigger than the neuron's passive time constant. In this work we showed that the interplay between $G_h^{Der}$ and $\tau_h$ expands the possibilities of existence of resonance beyond the restriction imposed by Hutcheon and Yarom. Since resonance can exist when $\tau_h \rightarrow 0$ if simultaneously $G_h^{Der} \rightarrow \infty$, then in principle resonance can exist even when $\tau_h < \tau_m$. 

Our phase plane analysis in the $A_h$-V plane allowed us to conclude that resonance emerges when the initial variation of $A_h$ is big but rapidly decreases when $\omega$ increases. This behavior is controlled by the interplay of two factors: $G_h^{Der}$, which is directly proportional to $dA^\infty_h/dV$ and determines the initial amplitude of $\Delta A_h$, and $\tau_h$, which determines the speed with which $\Delta A_h$ decreases when $\omega$ increases.

We also found that increasing $G_h^{Der}$ increases the resonance frequency while increasing $\tau_h$ decreases the resonance frequency. This is in agreement with experimental studies: whereas neurons from the auditory brainstem with high $G_K^{Der}$ and low $\tau_K$ display resonance with high frequency values ($f_{res} = 260$ Hz), CA1 pyramidal cells with low $G_h^{Der}$ and high $\tau_h$ (i.e. $G_h^{Der} < G_K^{Der}$ and $\tau_h > \tau_K$)  display resonance with low frequency values ($f_{res} = 2-7$ Hz) \cite{Mikiel2016,Hu2002}.

The resonance frequency is influenced by changes in $g_L$. Our results show that increasing $g_L$ ten-fold increases $\omega_{res}$. Our results suggest that in CA1 pyramidal cells $\omega_{res}$ reflect mainly the biophysical properties of the $I_h$ current, namely $G_h^{Der}$ and $\tau_h$, as reported elsewhere \cite{dougherty2013}. In addition, our results about the effect of $G_h^{Der}$, $\tau_h$ and $g_L$ are in agreement with previous experimental, computational and theoretical studies \cite{rotstein2014, hutcheon1996a, hutcheon1996b}. 

  When we studied the case of addition of an instantaneous current to the neuron model with leakage and $I_h$ currents, we observed an increase or decrease of the leak conductance for positive or negative values of the slope conductance ($G_i$), respectively. Clearly, this leads to an increase or decrease of the impedance magnitude and, finally, to a resonance amplification or attenuation \cite{ceballos2017,ceballos2017review,hutcheon1996a, hutcheon1996b}. However, this also leads to a novel effect since any change in the leak conductance is able to slightly change the probability of existence of resonance and the corresponding resonance frequency. Thus, we predict that $I_{NaP}$ and $I_{KIR}$ are able to slightly modulate the probability of existence of resonance and change the resonance frequency in neurons with $I_h$, e.g., CA1 pyramidal cells \cite{hutcheon1996a, hutcheon1996b,dougherty2013,rotstein2014}.  
  
In addition, given that the $I_h$ current is spread along the dendritic tree and soma with different kinetics, our results can be generalized for resonance properties in dendritic compartments at different positions of a neuron's dendritic tree. It has been shown that resonance properties in the different locations are related to intrinsic properties \cite{zhuchkova2013,landauski2014}. Moreover, different resonant currents also can be expressed at varying locations along the dendritic arbor with distinct dynamics. Since the dynamics for some other resonant currents can be defined using the same mathematical representation as the one adopted here for the $I_h$ \cite{gerstner2014}, our results can be adapted to study the resonance properties caused by these other currents. However, a study of the combined effect of many resonant currents distributed along a neuron's dendritic tree can only be done via computer simulations.

Our results have important implications for experimental research. For instance, we observed that for low $\tau_h$ values the resonance mainly exists for membrane potentials close to the maximum value of $G_h^{Der}$. This means that if there is no resonance for the membrane potential value where $G_h^{Der}$ is maximum, then we expect that there is no resonance for any other value of the membrane potential. In addition, the resonance frequency is highest for the membrane potential where $G_h^{Der}$ is maximum. Since it is experimentally challenging to measure resonance peaks for low frequencies, we predict that the best membrane potential candidate for the detection of a resonance by $I_h$ is the membrane potential where $G_h^{Der}$ is maximal. 

Furthermore, subthreshold resonance properties of neurons might be related to their information processing capabilities as they can influence the spiking characteristics of a neuron. For instance, the neuron's spiking response can follow the same frequency filtering selectivity of its subthreshold resonance \cite{richardson2003, blankenburg2015}, and this may influence network behavior \cite{chen2016}. Our results suggest a possible link between a measurable intrinsic neuronal property, namely its derivative conductance, and the spike times, which are usually considered as a basis for estimating the neuron's information processing capacity~\cite{strong1998,livroRieke}.

Consistent with previous studies \cite{hutcheon1996a,hutcheon1996b}, our results show that the main effect of $I_{h}$ is to attenuate the impedance at low frequencies and that decreasing $\tau_h$ enhances this attenuation. Unexpectedly, we also found that $I_{h}$ amplifies the impedance at high frequencies.

Previous studies have used a theoretical approach to elucidate the mechanisms by which $I_{h}$ generates resonance in neurons \cite{rotstein2014, richardson2003}. It has been challenging to relate these studies with experimental results. In order to fill this gap, we used a biophysical approach to study the mechanisms by which $I_{h}$ generates resonance in neurons. Our approach is testable by means of the derivative conductance $G_h^{Der}$ and the activation time constant $\tau_h$, which have been experimentally recorded as reported in \cite{dougherty2013,ceballos2017}. Therefore, we propose a verifiable hypothesis that the interplay between the derivative conductance and the activation time constant is the main biophysical mechanism underlying resonance in neurons.

\section{\label{sec:acknowledgment}Acknowledgment}

RFOP and CCC contributed equally to this work. This paper was developed within the scope of the IRTG 1740 / TRP 2015/50122-0, funded by DFG / FAPESP. This work was partially supported by the Research, Innovation and Dissemination Center for Neuromathematics (FAPESP grant 2013/07699-0). RFOP is supported by a FAPESP PhD scholarship (grant 2013/25667-8), CCC is supported by a CAPES PhD scholarship, VL is supported by a FAPESP MSc scholarship (grant 2017/05874-0), and ACR is partially supported by a CNPq fellowship (grant 306251/2014-0). Fruitful discussions with O. Kinouchi are greatly acknowledged.

\end{document}